\documentstyle[12pt]{article}
\begin{document}
\newcommand{\be}{\begin{equation}}
\newcommand{\ee}{\end{equation}}        
\newcommand{\w}{wavelet}
\newcommand{\an}{analysis}
\begin{center}
 Wavelets on nuclear interactions and QCD on jet multiplicities

\vspace{2mm}

 I.M. Dremin

\vspace{2mm}

{\it Lebedev Physical Institute, Moscow 117924, Russia}

\end{center}

\begin{abstract}
This talk at Fradkin conference is devoted to brief description of latest
results in two topics I worked on during last years.

The multiresolution analysis and fast wavelet transform became a standard 
procedure for pattern recognition and I describe how it has been used for
\an\ of high energy nucleus-nucleus interactions\footnote{Other applications
can be found, e.g., in Web site www.awavelet.com.}.

Latest QCD results on multiplicities in quark and gluon jets are discussed
and confronted to experimental data\footnote{Detailed review of this subject
is given in \cite{dg00}}.

\end{abstract}

\section{Introduction}
This talk is given at Fradkin memorial conference on Friday, 9 June 2000.
Friday was always a traditional day of seminars on quantum field theory in
our department, so-called Fradkin seminars. Two talks on wavelets were given
at this seminar last year just before Fradkin was put in a hospital. Therefore
I decided to talk here about wavelet application in particle physics and gave
to organizers of the conference the title containing the first part of the above 
one. However soon I learned that the topic on QCD (the second part of the 
present title) would fit the conference schedule better and asked organizers
for replacement. It was shifted to QCD session but with an old title on
wavelets. Thus I decided to give two talks at one, and it explains how these
two topics appear here together.

\section{Wavelets}

First I learned about wavelets from Pete Carruthers in 1993. He applied
them \cite{carr, lgca, gglc} for \an\ of some scaling cascade models used, 
in particular, in multiparticle production modelling. It was briefly described
in our review paper \cite{ddki}. Then I proposed to use wavelets for pattern
recognition in high energy nucleus-nucleus collisions, and it was applied
to experimental data \cite{adko}.

Wavelets became a powerful mathematical tool in many investigations. They
are used in those cases when the result of the analysis of a particular
signal\footnote{The notion of a signal is used here for any recorded 
information about some processes, objects, functions etc.} should contain 
not only the list of its typical
frequencies (scales) but also the knowledge of the definite local coordinates
where these properties are important. Wavelets form a complete orthonormalized
system of functions with a finite support by using dilations and translations.
 That is why by changing a scale (dilations) they can describe the local 
characteristics of a signal, and by translations they cover the whole region 
in which it is studied. Due to the completeness of the system, they also
allow for the inverse transformation to be done. The locality property of 
wavelets leads to their substantial superiority over Fourier transform which
provides us only with the knowledge of global frequencies (scales) of the object
under investigation because the system of functions used (sine, cosine) is 
defined on the infinite interval.

High energy collisions of elementary particles result in production of many 
new particles in a single event. Each newly created particle is depicted
kinematically by its momentum vector i.e. by a dot in the three-dimensional
phase space. Different patterns formed by these dots in the phase space
would correspond to different dynamics. To understand this dynamics is a 
main goal of all studies done at accelerators and in cosmic rays. Especially
intriguing is a problem of the quark-gluon plasma, the state of matter with
deconfined quarks and gluons which could exist during an extremely short 
intervals of time. One hopes to create it in collisions of high energy nuclei.
Nowadays, the data about Pb-Pb collisions are available where in a single event
more than 1000 particles are produced. We are waiting for RHIC accelerator in
Brookhaven and LHC in CERN to provide events with up to 20000 new particles 
created. Therefore the problem of phase space pattern recognition in an 
event-by-event \an\ becomes meaningful. It is believed that  the detailed 
characterization of each collision event could reveal the rare new phenomena,
and it will be statistically reliable due to a large number of particles 
produced in a single event.

When individual events are imaged visually, the human eye has a tendency
to observe different kinds of intricate patterns with dense clusters (spikes) 
and rarefied voids. However, the observed 
effects are often dominated by statistical fluctuations. The method of 
factorial moments was proposed \cite{bpes} to remove them but it is hard to use
in event-by-event approach. The \w\ \an\ avoids smooth polynomial trends 
typical for the statistical component. It was first applied \cite{adko} to 
analyze the individual high multiplicity event of Pb-Pb interaction at energy
158 GeV per nucleon. With emulsion technique used in experiment the angles
of particle emission are often measured only, and the two-dimensional phase
space is considered therefore. The experimental statistics is rather low but 
acceptance is high and homogeneous that is important for proper pattern 
recognition. To simplify the \an\ , the two-dimensional target diagram representing the
polar and azimuthal angles of created charged particles was split into 24
one-dimensional functions representing the polar angle distribution of 
these particles in 24 azimuthal angle sectors of $\pi /12$ and in each of them
particles were projected onto the polar angle $\theta $ axis. Thus 
one-dimensional functions of the rapidity distribution of these particles in 
24 sectors were obtained. Then the \w\ coefficients
were calculated in all of them and tied up together (continuous MHAT \w\
was used). The resulting pattern showed that many particles are concentrated
close to some value of the polar angle i.e. reveal the ring-like structure 
in the target diagram. The interest to such patterns is related to the fact
that they can result from the so-called gluon Cherenkov radiation 
\cite{dre1, dre2} 
or, more generally, from the gluon bremsstrahlung at a finite length within a
quark-gluon medium (plasma, in particular). More elaborated two-dimensional 
\an\ was done recently \cite{dikk} and confirmed these conclusions with
jet regions tending to lie on some ring-like formations. The jet-like
substructure of the event becomes more pronounced, and ring-like correlations 
of jettty regions are noticeable. With higher statistics, one can learn if the 
angular distribution of these rings corresponds to theoretical expectations.
It is due to \w\ \an\ 
that for the first time the fluctuation structure of an event is shown in a way
similar to the target diagram on the two-dimensional plot. 

Previously, some attempts \cite{addk, aaaa, cddh} to consider such events with 
different methods of treating the traditional projection and correlation 
measures revealed just that such substructures lead to spikes in
the angular (pseudorapidity) distribution and are somewhat jetty. Various Monte 
Carlo simulations of the process were compared to the data and failed to 
describe this jettiness in its full strength. More careful \an\ \cite{dlln, agab}
of large statistics data on hadron-hadron interactions
(unfortunately, however, for rather low multiplicity) with
dense groups of particles separated showed some "anomaly" in the angular 
distribution of these groups awaited from the theoretical side. Further \an\
using the results of \w\ transform are needed to check this conclusion in 
high multiplicity nucleus-nucleus interactions when many events of this kind
become available. 

\section{QCD}

Here I briefly describe recent advances in theoretical understanding of
multiplicity distributions of quark and gluon jets. The extended survey 
with more detailed comparison to experimental data can
be found in the recent review paper \cite{dg00}.

The progress in experimental studies of properties of quark and gluon jets is
very impressive. Therefore the study of
the energy evolution of such parameters of multiplicity distributions of jets
as their average multiplicities and widths becomes possible. It is well known
that the average multiplicities of quark and gluon jets increase quite fast
with energy but their ratio has a much slower dependence. 

The perturbative QCD provides quite definite
predictions which can be confronted to experiment. In brief, the results can be 
summarized by saying that the energy dependence of the mean jet
multiplicity can be perfectly fitted but the ratio of gluon to quark jet
multiplicities can be described within the precision of 15-20$\%$ only.
Moreover, one can understand why next-to-leading approximation is good enough
for describing the energy dependence, but it is not quite satisfactory yet for
the ratio value. I show this by presenting the analytical expressions. For the
corresponding Figures, I refer the reader to the review paper \cite{dg00}.

The theoretical asymptotical value of the ratio of
average multiplicities equal 2.25 is much higher than its
experimental values, which are in the range from 1.05 at 
comparatively low energies of $\Upsilon $ resonance to 1.5 at $Z^0$ resonance. 
The next-to leading order (NLO) corrections reduce this ratio from its
asymptotical value by about 10$\%$ at $Z^0$ energy. The NNLO and 3NLO
terms diminish it further and show the tendency to approximate the data 
with better accuracy.
The computer solution of QCD equations for the generating functions has shown 
even better agreement with experiment not only on this ratio but on
higher moments of multiplicity distributions as well. Being perfect
at $Z^0$ energy, the agreement in the ratio is not as good at lower energies
where the theoretical curve is still about 15-20$\%$
above the experimental one. In other words, the theoretically predicted 
{\em slope of the ratio} of multiplicities in gluon and quark jets is
noticeably smaller than its experimental value. Nevertheless, one can speak 
about the steady convergence of theory and experiment with subsequent 
improvements being done. Moreover, it is even surprising that any agreement
is achieved in view of the expansion parameter being extremely large
(about 0.5) at present energies.

The importance of studying the slopes stems from the fact that some of them 
are extremely sensitive (while others are not) to higher order
perturbative corrections and to non-perturbative terms in the available energy
region. Thus they provide us with a good chance to learn more
about the structure of the perturbation series from experiment. 

In the perturbative QCD, the general approach to studying the multiplicity
distributions is formulated in the framework of equations for generating
functions. Therefrom, one can get equations
for average multiplicities and, in general, for any moment of the multiplicity
distributions \cite{dr3}. In particular, two equations for average
multiplicities of gluon and quark jets are written as
\begin{eqnarray}
\langle n_G(y)\rangle ^{'} =\int dx\gamma _{0}^{2}[K_{G}^{G}(x)
(\langle n_G(y+\ln x)\rangle +\langle n_G(y+\ln (1-x)\rangle -\langle n_G(y)
\rangle ) \nonumber  \\
+n_{f}K_{G}^{F}(x)(\langle n_F(y+\ln x)\rangle +\langle n_F(y+
\ln (1-x)\rangle -\langle n_G(y)\rangle )],  \label{ng}
\end{eqnarray}
\begin{equation}
\langle n_F(y)\rangle ^{'} =\int dx\gamma _{0}^{2}K_{F}^{G}(x)
(\langle n_G(y+\ln x)\rangle +\langle n_F(y+\ln (1-x)\rangle -\langle n_F(y)
\rangle ).   \label{nq}
\end{equation}
Herefrom one can learn about the energy evolution
of the ratio of multiplicities $r$ and of the QCD anomalous dimension $\gamma $
(the slope of the logarithm of average multiplicity in a gluon jet) defined as
\begin{equation}
r=\frac {\langle n_G\rangle }{\langle n_F\rangle }\; ,\;\;\;\;\; \;\;\;
\gamma =\frac {\langle n_G\rangle ^{'}}{\langle n_G\rangle }
=(\ln \langle n_G\rangle )^{'}\; .  \label{def}
\end{equation}
Here, prime denotes the derivative over the evolution parameter
$y=\ln (p\Theta /Q_{0}),\\ 
p, \,\Theta $ are the momentum and the initial angular
spread of the jet, related to the parton virtuality $Q=p\Theta /2$,
\, $Q_{0}$=const, \, $K$'s are the well known splitting
functions, $\langle n_G\rangle $ and
$\langle n_F\rangle $ are the average multiplicities in gluon and quark jets, \,
$\langle n_G\rangle ^{'}$ is the slope of $\langle n_G\rangle $, \, $n_f$ is
the number of active flavours. The perturbative expansion of $\gamma $ and 
$r$ is written as
\begin{equation}
\gamma =\gamma _{0}(1-a_{1}\gamma _0 -a_{2}\gamma _{0}^{2} -a_{3}\gamma _{0}^{3}
)+O(\gamma _{0}^{5}),  \label{gam}
\end{equation}
\begin{equation}
r=r_0(1-r_{1}\gamma _{0}-r_{2}\gamma _{0}^{2}-r_{3}\gamma _{0}^{3})+O(\gamma 
_{0}^{4}),   \label{rat}
\end{equation}
where $\gamma _{0}=\sqrt {2N_{c}\alpha _{S}/\pi }, \, \alpha _{S}$
is the strong coupling constant, 
\begin{equation}
\alpha_{S}=\frac {2\pi }{\beta _{0}y}\left [1-\frac {\beta _{1}\ln (2y)}
{\beta _{0}^{2}y}\right ]+O(y^{-3}),   \label{alp}
\end{equation}
$\beta _{0}=(11N_{c}-2n_{f})/3, \, \beta _{1}=(51N_{c}-19 n_{f})/3, 
r_0 = N_c/C_F,$ \,and in QCD $N_{c}=3$ is the number of colours, $C_{F}=4/3$. 

The limits of integration in eqs. (\ref{ng}), (\ref{nq}) used to be chosen 
equal either to 0 and 1 or to $e^{-y}$ and $1-e^{-y}$. 
This difference, being negligibly small at high energies $y$, is quite
important at low energies. Moreover, it is of physics significance. With limits
equal to $e^{-y}$ and $1-e^{-y}$, the partonic cascade terminates at the 
perturbative level $Q_0$ as is seen from the arguments of multiplicities in the 
integrals. With limits equal to 0 and 1, one extends the cascade into the 
non-perturbative region with low virtualities $Q_{1}\approx xp\Theta /2$ and
$Q_{2}\approx (1-x)p\Theta /2$ less than $Q_{0}/2$. This region contributes
terms of the order of $e^{-y}$, power-suppressed in energy. It is not clear 
whether the equations and LPHD hypothesis are valid down to some $Q_0$
only or the non-perturbative region can be included as well.

Nevertheless, the purely perturbative expansion (\ref{gam}), (\ref{rat})
with constant coefficients $a_{i}, r_{i}$ and energy-dependent $\gamma _0$
is at work just in the case of limits 0 and 1. 
The values of $a_i , r_i$ for different number of
active flavors $n_f$ are tabulated in \cite{cdgnt}. 
At $Z^0$-energy the subsequent terms in (\ref{rat})
diminish the value of $r$ compared with its asymptotics $r_0 =2.25$
approximately by $10\%,\, 13\%,\, 1\%$ for $n_f=4$ getting closer to
experiment. However the theoretical
value of $r$ still exceeds its experimental values by 15-20$\%$.

The energy dependence of mean multiplicities can be obtained \cite{dgpl, cdgnt}
from the definition (\ref{def}) by inserting there the value of $\gamma $
(\ref{gam}) and integrating over $y$. Keeping the terms as small as $y^{-1}$
at large $y$ in the exponent, one gets \cite{cdgnt} the following expressions
for energy dependence of multiplicities of gluon (G) and quark (F) jets
\begin{equation}
\langle n_G\rangle =Ky^{-a_1C^2}\exp 
   \left[ 2C\sqrt y +\delta _G(y) \right], \label{ngy}
\end{equation}
with $K$ an overall normalization constant, $C=\sqrt {4N_c/\beta _0}$, and
\begin{equation}
\delta _G(y)=\frac {C}{\sqrt y}\left [ 2a_2C^2+\frac {\beta _1}{\beta _0^2}
[\ln (2y)+2]\right ] 
+\frac {C^2}{y}\left [ a_3C^2-\frac {a_1\beta _1}{\beta _0^2}
[\ln (2y)+1]\right ];   \label{del}
\end{equation}
\begin{equation}
  \langle n_F\rangle =\frac {K}{r_0}y^{-a_1C^2}\exp 
     \left[ 2C\sqrt y +\delta _F(y)\right], 
\label{nfy}
\end{equation}
with
\begin{equation}
\delta _F(y)=\delta _G(y)+\frac {C}{\sqrt y}r_1+\frac 
  {C^2}{y}(r_2+\frac {r_1^2}
  {2}).   \label{dfy}
\end{equation}
It happens that 2NLO and 3NLO terms (contributing to $y^{-1/2}$ and $y^{-1}$ 
terms in the exponent)
are almost constant at present energies and do not change the energy dependence
prescribed in NLO approximation. It explains why the energy dependence is well 
fitted by both NLO and 3NLO formulas while  2NLO correction to the ratio $r$ is large and
important. 

The rather small difference in $r$ values results
in quite noticeable disagreement of the slopes $r'$. 
Theoretical estimates can be shown \cite{cdgnt} to be quite predictive for 
the ratio of the slopes of multiplicities but it is much less
reliable to use the perturbative estimates even at $Z^0$-energy for such
quantities as the slope of $r$ or the ratio of slopes of logarithms of 
multiplicities (the logarithmic slopes). Much higher 
energies are needed to do that. Thus the values of $r^{'}$ and/or of
the logarithmic slopes can be used
to verify the structure of the perturbative expansion.

I demonstrate it here on the example of the slope value.
The slope $r'$ is extremely sensitive to higher 
order perturbative corrections. The role of higher order corrections is 
increased here compared with $r$ because each $n$th order term proportional
to $\gamma _{0}^{n}$ gets an additional factor $n$ in front of it when
differentiated, the main constant term disappears and the large ratio
$r_2/r_1$ becomes crucial:
\begin{equation}
r^{'} =Br_{0}r_{1}\gamma _{0}^{3}\left [1+\frac {2r_{2}\gamma _{0}}
{r_1}+\left (\frac {3r_3}{r_1}+B_{1}\right )\gamma _{0}^{2}+O(\gamma _{0}^{3})
\right ],       \label{rpri}
\end{equation}
where  the relation $\gamma _{0}^{'}\approx -B\gamma _{0}^{3}(1+B_{1}\gamma 
_{0}^{2})$ has been used with $B=\beta _{0}/8N_c ;\, 
B_{1}=\beta _{1}/4N_c\beta _0$. The factor in front of the bracket is very small
already at present energies: $Br_0r_1\approx 0.156$ and $\gamma _0\approx 0.5$.
However, the numerical estimate of $r^{'}$ is still indefinite due to the 
expression inside the brackets.
Let us note that each differentiation leads to a factor $\alpha _S$ or
$\gamma _{0}^{2}$, i.e., to terms of higher order.
For values of $r_1$, $r_2$, $r_3$ tabulated above ($n_f=4$) one estimates
$2r_{2}/r_{1}\approx 4.9$, $(3r_3/r_1)+B_1\approx 1.5$).
The simplest correction proportional to $\gamma _0$ is more than twice larger 1 
at energies studied and the next one is about 0.4. Therefore
the ever higher order terms should be calculated for the perturbative values
of $r'$ to be trusted. 
The slope $r'$ is equal to 0 for a fixed coupling constant.

The higher order terms are also important for the moments of the 
multiplicity distributions \cite{dln}.
The normalized second factorial moment $F_2$ defines the width of the
multiplicity distribution. 

The asymptotical $(\gamma _{0}\rightarrow 0)$ 
values of $F_{2}^{G}$ and $F_{2}^{F}$ are different:
\begin{equation}
F_{2, as}^{G}=\frac {4}{3}, \;\;\;\; F_{2, as}^{F}=\frac {7}{4}.  \label{fas}
\end{equation}
At $Z^0$ energy the widths of the distributions are smaller
\begin{equation}
F_{2}^{G}\approx 1.12, \;\;\;\; F_{2}^{F}\approx 1.34.   \label{fnum}
\end{equation}
but still much larger than their experimental values 1.02 and 1.08,
correspondingly. 
The rather large difference of the perturbative (\ref{fnum}) and experimental 
values at $Z^0$ indicates that moments of the distributions should be
sensitive to corrections. 
The conclusions about the third moments are similar.
Nonetheless, the computer solution of QCD equation happened to be quite
successful in fitting experimental data even for higher moments and their 
ratios $H_q$ introduced in \cite{dr1}. It shows that the role of conservation 
laws treated approximately in the analytical approach and accurately accounted 
in computer calculations becomes more important for higher moments.

Thus it is shown that the analytical approach is quite successful in
demonstrating that all features of QCD predictions about multiplicities of
quark and gluon jets correspond to the general trends of experimental data.
Some disagreement at the level of 15-20$\%$ is understandable due to 
incomplete account for the energy-momentum conservation in such an approach.
Further accurate computer solutions are needed to check if these trends
persist at the higher precision level.

\end{document}